\documentclass[12pt]{JHEP3}
\usepackage{mathtools}

\usepackage{amsmath,epsfig}
\usepackage{amssymb,amsfonts}
\usepackage{latexsym}
\usepackage[latin1]{inputenc}
\usepackage{xcolor}
\let\normalcolor\relax
\usepackage{graphicx}
\usepackage{url}
\usepackage{slashed}
\usepackage{tikz}

\usetikzlibrary{patterns}

\usepackage{longtable}

\relax
\renewcommand{\theequation}{\arabic{section}.\arabic{equation}}
\def\be{\begin{equation}}
\def\ee{\end{equation}}
\def\bea{\begin{eqnarray}}
\def\eea{\end{eqnarray}}

\newcommand\fverb{\setbox\pippobox=\hbox\bgroup\verb}
\newcommand\fverbdo{\egroup\medskip\noindent%
                        \fbox{\unhbox\pippobox}\ }
\newcommand\fverbit{\egroup\item[\fbox{\unhbox\pippobox}]}

\newcommand{\bear}{\begin{eqnarray}}

\newcommand{\eear}{\end{eqnarray}}

\newcommand{\bsea}{\begin{subeqnarray}}
\newcommand{\esea}{\end{subeqnarray}}
\newbox\pippobox

\def\6{\partial}

\def\a{\alpha}

\def\sq
\def\a{\alpha}

\def\hri#1#2{\href{http://arxiv.org/abs/#1}{[ArXiv:#1]#2}}
\def\hre#1#2{\href{http://arxiv.org/abs/#1/#2}{[ArXiv:#1/#2]}}

\newcommand{\eq}[1]{(\ref{#1})}

\title{Gravitational collapse and thermalization in the hard wall model}
\author{\Large Ben Craps$^a$, Elias Kiritsis$^{b,c,d}$, Christopher Rosen$^d$, Anastasios Taliotis$^a$, Joris Vanhoof$^a$, Hongbao Zhang$^a$ \\
~\\$^a$ \href{http://we.vub.ac.be/tena/}{Theoretische Natuurkunde, Vrije Universiteit Brussel}, and \\
\hspace*{0.15cm}  \href{http://www.solvayinstitutes.be/}{International Solvay Institutes},
Pleinlaan 2, B-1050 Brussels, Belgium\\
~\\
$^b$ \href{http://www.apc.univ-paris7.fr}{APC, Universit\'e Paris 7, Diderot},
 CNRS/IN2P3, CEA/IRFU, Obs. de Paris, Sorbonne Paris Cit\'e,
  B\^atiment Condorcet, F-75205, Paris Cedex 13, France (UMR du CNRS 7164)\\
~
\\
$^c$\href{http://wwwth.cern.ch/}{Theory Group, Physics Department, CERN}, CH-1211, Geneva 23, Switzerland\\
~\\
$^d$ \href{http://hep.physics.uoc.gr}{Crete Center for Theoretical Physics},
Department of Physics, University of Crete, 71003 Heraklion, Greece.
\\\\
E-mail: Ben.Craps@vub.ac.be, kiritsis@physics.uoc.gr, rosen@physics.uoc.gr, 
Anastasios.Taliotis@vub.ac.be, Joris.Vanhoof@vub.ac.be, hzhang@vub.ac.be
}

\preprint{CCTP-2013-20\\CCQCN-2013-07\\CERN-PH-TH/2013-267}

\abstract{We study a simple example of holographic thermalization in a confining field theory: the homogeneous injection of energy in the hard wall model. Working in an amplitude expansion, we find black brane formation for sufficiently fast energy injection and a scattering wave solution for sufficiently slow injection. We comment on our expectations for more sophisticated holographic QCD models.}


\begin{document}

\section{Introduction}

The use of gauge/gravity duality to study the thermalization of strongly coupled field theories has been an active area of research, with potential applications to ultrarelativistic heavy ion collisions, strongly correlated electron systems and cold atoms. In particular, heavy ion collisions at RHIC or LHC are often modeled as the sudden injection of energy in a conformal field theory with gravity dual, or as the collision of sheets of energy in such a theory.

QCD is not a conformal theory, but conformal invariance is in any case broken in the resulting finite temperature state. Therefore, one may hope that the simplest AdS/CFT models will share qualitative features with QCD, perhaps even allowing order-of-magnitude estimates for certain quantities, such as thermalization times. Having said this, it would clearly be interesting to study thermalization in holographic theories that are closer to QCD, in particular in confining theories. The purpose of the present paper is to do so for the simplest model, namely the homogeneous injection of energy in the hard wall model  \cite{ps}.\footnote{Shock wave collisions in confining models have been studied in \cite{Kiritsis:2011yn,Cardoso:2013vpa}.}

Generic holographic thermalization models require the use of numerical general relativity \cite{Chesler:2013lia}, and it will be interesting to extend those techniques to confining models. However, interesting models exist that allow an analytic treatment \cite{BM}, and a confining version of those will be the main focus of the present paper. Starting from the field theory vacuum, we will briefly turn on a source with amplitude of order $\epsilon\ll 1$ during a time $\delta t$. This causes a shell of energy to fall into the interior of the dual bulk spacetime. In the case that the boundary field theory is a CFT$_{d}$ (dual to AdS$_{d+1}$), this process is described in \cite{BM}.\footnote{See \cite{Wu:2012rib} for a numerical study of this process (see also \cite{Bizon:2011gg, Jalmuzna:2011qw, Garfinkle:2011hm, Garfinkle:2011tc} for similar techniques used in a slightly different setup), and \cite{Chesler:2008hg} for earlier work on a closely related setup.} For a translationally invariant setup\footnote{Inhomogeneities were included in \cite{Balasubramanian:2013rva}.}, it was shown that this always results in black brane formation at small amplitude (which physically corresponds to the injection time being short compared to the inverse temperature of the black brane to be formed). To leading non-trivial order in the amplitude $\epsilon$ the black brane horizon radius is given by $r_{h}\sim\epsilon^{2/d}/\delta t$. To this order, the bulk geometry is given by the AdS-Vaidya metric, which has turned out to be a very useful model for holographic thermalization. In various works, the time evolution of various probes has been computed, including expectation values of local gauge-invariant operators \cite{BM}, spectral functions \cite{Balasubramanian:2012tu}, equal-time two-point functions, Wilson loops \cite{Balasubramanian:2010ce}, entanglement entropy \cite{Balasubramanian:2010ce,Hubeny:2007xt, AbajoArrastia:2010yt,Albash:2010mv, Liu:2013iza}, mutual and tripartite information \cite{Balasubramanian:2011at,Allais:2011ys,Callan:2012ip},  and causal holographic information \cite{Hubeny:2013hz}.\footnote{See \cite{Balasubramanian:2012tu} for a longer list of references to related work.}

The paper \cite{BM} also analyzed spherical shell collapse in {\em global} AdS, dual to the homogeneous injection of energy on a sphere at some moment of time. Here, a much richer structure was found, depending on the ratio $x\equiv\delta t/R$, with $R$ the radius of the sphere, and we focus on $x\ll 1$. For $x\ll\epsilon^{2/d}$,  one forms a large black hole; for $\epsilon^{2/d}\ll x\ll \epsilon^{1/(d-1)}$ a small black hole; and for $\epsilon^{1/(d-1)} \ll x$ a wave that scatters back to the boundary of AdS (after which it would reflect from the boundary, leading to more complicated evolution on longer timescales, the detailed analysis of which was beyond the scope of \cite{BM}).\footnote{The question whether global AdS is nonlinearly stable under perturbations similar to those considered in \cite{BM} has been the subject of recent debate. In \cite{Bizon:2011gg}, it was argued that a weakly turbulent instability leads to black hole formation after a number of reflections from the boundary. More recent work \cite{Dias:2012tq,Maliborski:2013jca,Buchel:2013uba} has shown that this may or may not happen depending on the details of the initial energy distribution.} A review of these results can be found in Appendix~\ref{reviewBM}. 

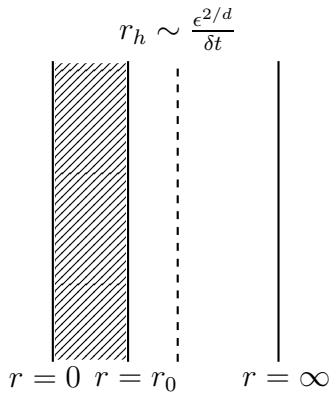
\begin{figure}[!h]
\centering
\begin{tikzpicture}
\draw[pattern=north east lines, pattern color=black] (0,0) rectangle (1,4);
\draw[ultra thick,color=white] (0,0) rectangle (1,4);
\draw[thick] (0,0) -- (0,4);
\draw (0,0.075) node [text=black,below] {$r=0\,\,\,$};
\draw[thick] (1,0) -- (1,4);
\draw (1,0) node [text=black,below] {$\,\,\, r=r_{0}$};
\draw[thick] (3,0) -- (3,4);
\draw (3,0) node [text=black,below] {$\,\,\, r=\infty$};
\draw[dashed,thick] (1.66,0) -- (1.66,4) node [text=black,above] {$r_{h}\sim\frac{\epsilon^{2/d}}{\delta t}$};
\end{tikzpicture}
\caption{\emph{Schematic representation of the bulk in the hard wall model. The portion $r<r_0$ is removed from the spacetime, so only black branes with $r_h>r_0$ can be formed.
}}\label{fig:hardwall}
\end{figure}

These possibilities depend crucially on the structure of black hole solutions in global AdS (namely on the existence of both large and small black holes). 
The starting point of our present paper is that a similarly rich structure of black brane solutions can be found in confining holographic models with the dual field theory living in Minkowski spacetime (as opposed to a cylinder as for global AdS). An overview of confining holographic models and their black brane solutions is given in section~\ref{sec:confining}. The simplest of these confining models is the hard wall model, which we will analyze using the techniques of \cite{BM}. In this model,
schematically depicted in figure \ref{fig:hardwall}, the background geometry is AdS$_{d+1}$, but the spacetime is cut off at some finite value $r=r_{0}$, which corresponds to the location of the so called hard wall. The radial coordinate $r$ then ranges from $r_{0}$ to $\infty$. The location of the hard wall is proportional to the confinement scale $\Lambda$ of the boundary theory: $r_{0}\simeq\Lambda$. This model only has black branes with event horizon $r_{h}\sim\frac{\epsilon^{2/d}}{\delta t}$ larger than $r_{0}$.

In the remainder of this introduction, we first review various confining holographic models and their black brane solutions, after which we motivate the use of the bulk axion to inject energy. Then we describe our results in the hard wall model. Finally, we discuss these results, and speculate on what would happen in other models.

\subsection{Confining holographic theories}
\label{sec:confining}


While the concrete results of the present paper will be limited to the simplest confining model, our eventual goal is to study more realistic holographic models for QCD. With that purpose in mind, we now give an overview of confining holographic models, with an emphasis on their finite temperature solutions (in particular black branes), which are crucial for studies of thermalization.

Confinement is defined in terms of the behavior of the expectation value of the Wilson loop operator in a given (semiclassical) state. In the holographic  context, the general conditions for confinement were studied in \cite{cobi}. The conclusion is that if the {\em string frame} metric scale factor has a minimum and at that minimum the scale factor is non-zero, then the Wilson loop shows area behavior, the string tension is finite and the theory confines.

There are several holographic setups that implement confinement in holography.
Among the top-down ones, the earliest and simplest is the black D4-brane model \cite{D4}, which describes 5d maximally supersymmetric Yang-Mills (YM) compactified to 4d on a circle with supersymmetry breaking boundary conditions. Its geometry involves a conformally flat Minkowski part, and a cigar geometry involving the holographic radial direction and the fifth (compact) direction.
In this solution there is confinement,  as there is an endpoint in the bulk geometry (the tip of the cigar) where the scale factor reaches a minimum while remaining finite in accordance  with \cite{cobi}.
There are top-down generalizations of this setup that have  more complicated geometries.
At finite temperature the black D4 theory has a non-trivial structure. There is a first order transition \cite{D4t}, to a deconfined phase, but there is also the analogue of the small black hole solutions at any temperature. Moreover, these solutions have vanishing temperature at vanishing horizon size and are therefore locally (perturbatively) stable. Moreover,  at high temperature  the physics is five-dimensional. The phase structure was recently challenged in \cite{MM}.

There are also bottom-up holographic theories that implement confinement and can be  tuned to describe YM theory at large $N$.
The crudest of all is the so called hard wall model \cite{ps}. It is a slice of $AdS_5$ where the radial coordinate varies from the boundary to an IR cutoff $r_0$. The presence of this IR ``wall" provides by fiat a minimum for the scale factor and again a computation of the Wilson loop here indicates confinement \cite{cobi}.
The hard wall background was used to describe, with some success, the meson sector \cite{adsqcd}, by providing a confining background on which the flavor fields propagate. It has also been rather successful in fitting deep inelastic scattering data from HERA \cite{Costa:2013uia,Costa:2012fw,Brower:2010wf}. Its main advantage is its simplicity, although it falls short in several ways in describing the dynamics of large-$N$ YM. At zero temperature, it exhibits confinement, but
\begin{itemize}
\item The glueball (radial) trajectories have masses that asymptote as $m_{n}\sim n$ for large $n$, instead of the expected $m_n\sim \sqrt{n}$.
\item The magnetic charges are confined (instead of screened) \cite{ihqcd1,ihqcd2}.
\end{itemize}
At finite temperature it exhibits a deconfining phase transition \cite{adst}, to a black brane phase. The black branes are of course the large AdS-Schwarzschild black holes and implicit in the transition is that such black branes exist only if the horizon position in the radial direction $r_h$ is outside the hard wall, $r_h>r_0$.
Moreover, the equation of state for the deconfined phase is exactly conformal.

A modification of the hard wall model, the soft-wall model \cite{soft}, was introduced in order to render meson radial trajectories linear. In the gluon sector, however, the background does not satisfy equations of motion and therefore thermodynamics is ill-defined among other things.%
\footnote{
By contrast, the hard wall model can be rigorously defined by putting a boundary in the infrared and imposing appropriate boundary conditions \cite{Randall:1999vf}. Then the solution is an extremum of a gravitational action, and thermodynamics makes sense. This cannot be done for the soft wall model \cite{}, and this is why the energy and the entropy computed in the soft wall model do not satisfy the first law. 
}

A more sophisticated bottom-up model for YM is Improved Holographic QCD (IHQCD) \cite{ihqcd1,ihqcd2}. It was conceived to abide by string theory input and holographic dictums, and at the same time match YM features at zero and finite temperature (a review can be found in \cite{rev} and a string theory motivation in \cite{dis}).
The non-trivial confining geometry in IHQCD is driven by a dilaton potential, which implements the renormalization group running of the YM coupling, dual to the bulk dilaton field. As the bulk theory is five-dimensional, there is a mild singularity in the IR end of the geometry that is repulsive\footnote{This in particular means that (a) it satisfies the Gubser bound \cite{gubser},  (b) Wilson loops remain always a finite distance from the singularity, (c) the fluctuation problem and associated spectra do not depend on the resolution of the singularity \cite{ihqcd2}.}  and therefore innocuous for low energy physics. The theory at zero temperature exhibits confinement, a mass gap, and linear (i.e., $m_n\sim \sqrt{n}$) glueball trajectories. It is also interesting that the feature of the dilaton potential responsible for the linear trajectories, is also responsible for the $T^2$ behavior of the free energy just above the deconfining transition \cite{gg}.
Confinement in IHQCD happens non-trivially. Although the Einstein frame scale factor is monotonic and decreasing towards the IR as the null energy condition dictates, the string frame scale factor is different (as the dilaton is non-trivial), and it has a non-trivial minimum at the interior of the geometry, and always at finite distance from the IR singularity. The existence of this minimum is responsible for the confining property \cite{ihqcd2}.

At finite temperature, and up to a minimum temperature $T_{\min}$ there are no black brane solutions. This property is directly correlated with the existence of the mass gap in the zero temperature theory \cite{gkmn1,gkmn2}.
Therefore, for $T<T_{min}$ there is a single saddle-point solution with the appropriate boundary conditions, namely the thermal-gas solution (this, as usual,  is the $T=0$ solution with time compactified in accordance with the temperature). The system therefore is in the confined phase.
At $T>T_{min}$ apart from the thermal gas solutions, there are also two black brane solutions, a small (and thermodynamically unstable) one, and a large one (which is stable). This situation is reminiscent of global AdS, although here the black brane horizons are flat. Finally, at $T_c>T_{min}$ there is a first order phase transition to the large black brane solution that models the deconfined plasma phase. The thermodynamics of small black branes in IHQCD is very different from those of flat space. It has been analyzed in \cite{KT} where the formation of black branes in heavy ion collisions was discussed.

By tuning two phenomenological parameters, IHQCD can describe very well both $T=0$ glueball spectra, as well as the finite temperature thermodynamic functions \cite{data}.
A recent high-precision lattice study of large-$N$ YM thermodynamics has indicated that $N=3$ is very close to $N=\infty$, and that it agrees very well with IHQCD \cite{panero}, (see \cite{LP} for a recent review of large-$N$ lattice calculations). Moreover, the model has been used to compute the bulk viscosity as well as heavy quark energy loss \cite{transport,langevin}.

There is an alternative model for describing the thermodynamics of QCD, due to Gubser and Nellore \cite{gn}. The focus of this model was to describe QCD thermodynamic functions well and in particular to have a rapid crossover in the entropy rather than a phase transition, a fact valid in QCD with small but finite masses for the light quarks. Therefore, it does not have confinement at zero temperature.
This affects the structure of the black brane solutions. The model has black branes at all temperatures and a ``deconfining" transition at $T=0^+$ to  the black brane phase. In this model, all black branes are stable. 

As we have seen above, different models that exhibit confinement have a different structure of black branes at finite temperature, and this signals that they will probably have different processes for thermalization. In this work, we focus on the simplest of them, the hard wall model.

\subsection{Glueballs and axions}

There is a further issue in the thermalization problem we are studying: the ``channel" we use to inject energy into the system. In the conformal case, a massless scalar was used in \cite{BM}, mainly because it is the simplest to describe.

In YM there are in principle several fields that can be used to inject energy. One is the stress-tensor, corresponding to injecting energy via the metric in the dual holographic theory \cite{Chesler:2008hg}. Another operator is the YM lagrangian, ${\rm Tr}[F^2]$, dual to the dilaton. Although this operator is marginal in the UV (and therefore dual to a massless bulk scalar), it acquires a non-trivial anomalous dimension, and becomes strongly relevant in the IR. It can be used to inject energy, and this is a very interesting way to do it, however the analytic problem is very hard.

YM has also another marginal operator, the instanton density, ${\rm Tr}[F\wedge F]$. It is dual to a bulk axion field. The fact that this operator is marginal to all orders in perturbation theory indicates that the dual scalar has no potential to leading order in the large-$N$ expansion\footnote{YM instantons will eventually generate a potential. The expectation is  that
it will be exponentially suppressed at large $N$.}. This is in accord with the string theory Peccei-Quinn shift symmetry of the string axion.
Of course, non-perturbative contributions will affect the renormalization of this operator, and these appear in terms of kinetic mixing between the dilaton and the axion \cite{data,CS}. The spectrum associated with this bulk field gives rise to the $0^{-+}$ glueball tower in YM theory.

In the hard wall model, there is no running of the dilaton and therefore the kinetic mixing is not relevant. We may therefore consider the axion as a true massless bulk field and this is what we will use in the sequel to inject energy into the theory.

\subsection{Results}

We inject energy in the hard wall model using a massless scalar field $a$. While many of our formulas will be valid in more general dimensions, we focus on the case where the boundary theory has $d=3$ space-time dimensions for simplicity and concreteness. As mentioned before, the energy scale $\Lambda = r_0$, defined by the position of the hard wall, defines the confinement scale of the theory. At a given time $t=0$ we briefly turn on a homogeneous source $a_{s}$ for (the operator dual to) the scalar field with amplitude of order $\epsilon$ during a time $\delta t$. Determining the result of this perturbation at the UV boundary in the bulk corresponds to solving the full set of Einstein's equations with the boundary condition for the scalar field determined by the source $a_{s}$. This is a system of coupled, non-linear partial differential equations. These equations can be linearised by expanding the fields in the amplitude $\epsilon\ll1$ of the disturbance and then solved order by order. For arbitrary profiles of the source $a_{s}$, we find explicit solutions for the leading corrections to the background in the amplitude expansion which are of order $\epsilon$ for the scalar field and of order $\epsilon^{2}$ for the metric. By analyzing these perturbative solutions, we can separate two clearly distinct classes of solutions depending on the amplitude $\epsilon$, the injection time $\delta t$ and the location of the hard wall $\Lambda$. For $d=3$ and $\Lambda\,\delta t\ll1$ we have the following two cases:\footnote{The complementary regime $\Lambda\,\delta t\gg1$ is discussed in Section \ref{Regime of validity of perturbation theory}.}

\begin{itemize}
\item If $\epsilon^{2}\gtrsim(\Lambda\,\delta t)^{3}$, an AdS-Schwarzschild black brane is formed in the bulk, with event horizon $r_{h}\sim\frac{\epsilon^{2/3}}{\delta t}$. The leading non-trivial terms in the $\epsilon$-expansion describe an infalling solution for the scalar field and a Vaidya type metric. As explained in \cite{BM}, naive perturbation theory in $\epsilon$ breaks down for times of order $1/T\sim 1/r_h$, but can be resummed by taking AdS Vaidya (rather than AdS) as the starting point for the perturbative expansion. The hard wall remains well inside the event horizon at all times, so it does not influence the solution outside the event horizon, and the process is essentially identical to that studied in \cite{BM}.
\item If $\epsilon^{2}\ll(\Lambda\,\delta t)^{5}$, the infalling shell scatters from the hard wall. To leading non-trivial order in the $\epsilon$-expansion, the shell keeps scattering between the hard wall and the UV boundary. The leading backreaction on the metric can be shown to remain small compared to the background. We comment on corrections to this picture in section~\ref{discussion}. This regime is analogous to the scattering wave regime in global AdS \cite{BM}, with the confinement scale playing the role of $1/R$. 
\end{itemize}

\begin{figure}[!h]
\centering
\begin{tikzpicture}
\draw[thick,->] (0,0) -- (0,3) node [text=black,left] {$\epsilon$};
\draw[thick,->] (0,0) -- (5,0) node [text=black,right] {$\Lambda\delta t$};
\draw[thick] plot [smooth, tension=1] coordinates {(0,0) (0.5,0.17) (1,0.47) (1.5,0.87) (2,1.34) (2.5,1.86) (3,2.46) (3.5,3.11)};
\draw[thick] plot [smooth, tension=1] coordinates {(0,0) (0.5,0.01) (1,0.05) (1.5,0.13) (2,0.27) (2.5,0.47) (3,0.74) (3.5,1.09) (4,1.52) (4.5,2.04) (5,2.65)};
\draw[fill,white] (3,2.65) rectangle (6,3.2);
\draw[thick] (2.7,3) node [text=black,right] {$\sim(\Lambda\delta t)^{\frac{3}{2}}$};
\draw[thick] (4.7,3) node [text=black,right] {$\sim(\Lambda\delta t)^{\frac{5}{2}}$};
\draw[thick] (1.3,2.1) node [text=black] {Black brane};
\draw[thick] (4.8,1) node [text=black] {Scattering};
\end{tikzpicture}
\caption{\emph{Different phases depending on the amplitude $\epsilon$, the injection time $\delta t$ and the location of the hard wall $\Lambda$.
}}\label{fig:TwoRegimes}
\end{figure}
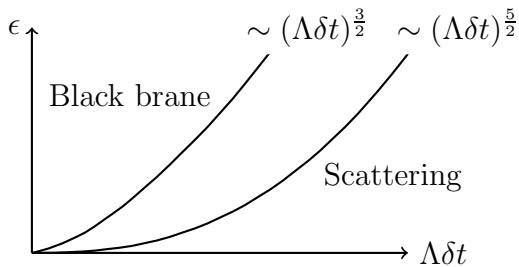

So as could have been expected, confinement drastically alters the thermalization process. In the absence of a hard wall ($\Lambda=0$), a black brane would always be formed in planar AdS. When a hard wall is present, two different ``phases'' can be distinguished (see Figure \ref{fig:TwoRegimes}). The perturbative analysis presented in this paper is insufficient to determine the bulk solution in the intermediate regime.

\subsection{Discussion and outlook}
\label{discussion}

We have seen that to leading non-trivial order in the amplitude expansion, for short injection times a black brane is formed, whereas for longer injection times (but still short compared to the QCD scale), the shell scatters back and forth between the boundary and the hard wall. In the regime where a large black brane is formed dynamically, we are automatically in the temperature range where the large black brane is thermodynamically favored (namely above the confinement scale $\Lambda$). 

An important question is how nonlinear effects modify the scattering solution. For an infalling shell in global AdS, weak turbulence may occur, causing energy to be focused and a small black hole to be formed after one or more reflections from the AdS boundary \cite{Bizon:2011gg} (see, however, \cite{Dias:2012tq,Maliborski:2013jca,Buchel:2013uba}). In the hard wall model, however, there are no obvious analogues of these small black holes\footnote{At least no translationally invariant ones.}, and the scattering solution does not contain enough energy to form an AdS-Schwarzschild black brane (since the black brane it would form in the absence of the hard wall would have a horizon ``inside'' the wall, meaning it would be lighter than any AdS-Schwarzschild black brane with horizon outside the wall).

This still leaves open the question what the final state of the ``scattering solution'' will be: does the solution keep oscillating forever? We expect that over a sufficiently long timescale, the solution will eventually thermalize (into a thermal gas, not a black brane), as was expected by \cite{BM} for global AdS (before the weakly turbulent instability of related models had been pointed out  \cite{Bizon:2011gg}).

An obvious next step is to understand the effects of a more realistic confining geometry on the thermalization of a scalar pulse. In confining models with small black branes, including IHQCD, we expect the situation to be more complicated, and more similar to that in global AdS. Depending on the injection time, we expect the formation of a large black brane, a small black brane or (at least initially) a scattering wave. It would be interesting to explore the late-time behavior of the scattering wave solution, which might exhibit a weakly turbulent instability towards the formation of a small black brane. We are currently carrying out this analysis, which requires numerical general relativity, and hope to report on it in a sequel to the present paper.  

Heavy ion collisions provide an obvious motivation for this work, so it is tempting to speculate on (naive) extrapolations of our results to QCD. For collisions at RHIC or the LHC, the crossing time $\delta t$ is short compared to the (initial) temperature $T$ of the plasma to be formed (which implies $\epsilon\sim (T\delta t)^{d/2} < 1$ and is therefore consistent with the amplitude expansion) and the QCD scale (which implies that we would either form a black brane or a thin scattering shell). Comparing $\Lambda \delta t$ with $\epsilon^{2/d}$, we can see that in the regime in which a black brane is formed, the temperature is high compared to the QCD scale, corresponding to a deconfined plasma.

\section{Setup of the model}

The gravitational dual of our model is determined by the Einstein-Hilbert action including a negative cosmological constant which is minimally coupled to a massless scalar field:
\begin{equation}
\mathcal{S}=\frac{1}{2\kappa^{2}}\int\text{d}^{d+1}\mathbf{x}\sqrt{-g}\left(R+\frac{d(d-1)}{L^{2}}-\frac{1}{2}(\partial a)^{2}\right).
\end{equation}
The above action results in the equations of motion
\begin{equation}
E_{\mu\nu}\equiv G_{\mu\nu}-\frac{d(d-1)}{2L^{2}}g_{\mu\nu}-\left(\frac{1}{2}\partial_{\mu}a\partial_{\nu}a-\frac{1}{4}g_{\mu\nu}(\partial a)^{2}\right)=0
\label{eq:Einstein}
\end{equation}
and
\begin{equation}
\Box a\equiv\frac{1}{\sqrt{-g}}\partial_{\mu}\left(\sqrt{-g}g^{\mu\nu}\partial_{\nu}a\right)=0.
\end{equation}

We are interested in a translationally invariant setup and will focus on homogeneous planar solutions to these equations. Therefore we use the following ansatz for the metric and the scalar field:
\begin{equation}
ds^{2}=-g(r,v)dv^{2}+2drdv+f^{2}(r,v)d\vec{x}^{2}
\qquad\text{and}\qquad
a=a(r,v).
\end{equation}
This form ensures that $E_{vx_{j}}=E_{rx_{j}}=0$ and $E_{x_{i}x_{j}}=0$ (for $i\neq j$) are automatically satisfied. The conditions $E_{rr}=0$, $E_{rv}=0$ and $\Box a=0$ imply that $E_{x_{i}x_{i}}=0$ and $\frac{d}{dr}\left(rE_{vv}\right)=0$. Therefore it is sufficient to solve the equations $E_{rr}=0$, $E_{rv}=0$ and $\Box a=0$, supplemented by imposing $E_{vv}=0$ at a particular value of $r$, to solve all components of equation (\ref{eq:Einstein}). Since
\begin{align*}
E_{rr}&=-(d-1)\frac{\partial_{r}^{2}f}{f}-\frac{1}{2}(\partial_{r}a)^{2}, \\
2E_{rv}+gE_{rr}&=\frac{(d-1)}{f^{d-1}}\partial_{r}\left(f^{d-2}\left(g\partial_{r}f+2\partial_{v}f\right)\right)-\frac{d(d-1)}{L^{2}}, \\
E_{vv}+gE_{rv}&=\frac{(d-1)}{2f}\left(-2\partial_{v}^{2}f+\partial_{v}f\partial_{r}g-\partial_{r}f\partial_{v}g-2g\partial_{v}\partial_{r}f\right) \\
&\qquad\qquad\qquad\qquad\qquad\qquad-\frac{1}{2}\left((\partial_{v}a)^{2}+g(\partial_{r}a)(\partial_{v}a)\right), \\
\Box a&=\frac{1}{f^{d-1}}\left(\partial_{r}\left(f^{d-1}g\partial_{r}a\right)+\partial_{r}\left(f^{d-1}\partial_{v}a\right)+\partial_{v}\left(f^{d-1}\partial_{r}a\right)\right),
\end{align*}
these requirements result in the equations
\begin{equation}
-2(d-1)\partial_{r}^{2}f=f(\partial_{r}a)^{2},
\label{eq:metric1}
\end{equation}
\begin{equation}
\partial_{r}\left(f^{d-2}\left(g\partial_{r}f+2\partial_{v}f\right)\right)=\frac{df^{d-1}}{L^{2}}
\label{eq:metric2}
\end{equation}
and
\begin{equation}
\partial_{r}\left(f^{d-1}g\partial_{r}a\right)+\partial_{r}\left(f^{d-1}\partial_{v}a\right)+\partial_{v}\left(f^{d-1}\partial_{r}a\right)=0.
\label{eq:scalar}
\end{equation}

In the absence of the scalar field ($a(r,v)=0$) these differential equations are solved by $f(r,v)=r/L$ and $g(r,v)=(r/L)^{2}$. The metric thus becomes
\begin{equation}
ds^{2}=-\frac{r^{2}}{L^{2}}dv^{2}+2drdv+\frac{r^{2}}{L^{2}}d\vec{x}^{2}=L^{2}\frac{dr^{2}}{r^{2}}+\frac{r^{2}}{L^{2}}(-dt^{2}+d\vec{x}^{2}),
\end{equation}
which is that of AdS$_{d+1}$ in Poincar\'e coordinates after the substitution $v=t-\frac{L^{2}}{r}$. By convention, we set the AdS radius $L$ in this text equal to 1, such that all quantities are dimensionless. Now we will perturb this background by turning on a small source $a_{s}(v)$ (of order $\epsilon$) at the boundary around $v=0$ during an injection time $\delta t$:
\begin{align}
a_{s}(v)=0 \hspace{0.95cm} & \quad (v<0) \nonumber \\
a_{s}(v)=\epsilon\,a_{0}(v) & \quad (0<v<\delta t) \\
a_{s}(v)=0 \hspace{0.95cm} & \quad (\delta t<v) \nonumber
\end{align}

To investigate the influence of this perturbation, we will solve equations (\ref{eq:metric1}, \ref{eq:metric2}, \ref{eq:scalar}) with initial conditions given by the background
\begin{equation}
f(r,v)=r
\quad\text{,}\quad
g(r,v)=r^{2}
\quad\text{and}\quad
a(r,v)=0
\qquad\text{for}\qquad
v<0.
\end{equation}
The boundary conditions at the UV boundary ($r\rightarrow\infty$) in our setup are given by
\begin{equation}
\lim_{r\rightarrow\infty}\frac{f(r,v)}{r}=1
\quad\text{,}\quad
\lim_{r\rightarrow\infty}\frac{g(r,v)}{r^{2}}=1
\quad\text{and}\quad
\lim_{r\rightarrow\infty}a(r,v)=a_{s}(v).
\end{equation}
In fact, in order to fix the gauge redundancy of our metric ansatz completely, we need to restrict these further to
\begin{align}
f(r,v)&=r\left(1+\mathcal{O}\left(\frac{1}{r^{2}}\right)\right), \\
g(r,v)&=r^{2}\left(1+\mathcal{O}\left(\frac{1}{r^{2}}\right)\right), \\
a(r,v)&=a_{s}(v)+\mathcal{O}\left(\frac{1}{r}\right).
\end{align}
This setup so far corresponds exactly to the one used in \cite{BM}. The additional boundary condition that we need to impose in our setup is due to the presence of a hard wall in the bulk. At the location of the hard wall ($r=r_{0}$) we impose Neumann boundary conditions on the scalar field
\begin{equation}\label{eqn:NeumannBoundaryCondition}
0=\left.(n^{\mu}\partial_{\mu})a(r,v)\right|_{r=r_{0}}=\left.\left(\sqrt{g}\frac{\partial}{\partial r}+\frac{1}{\sqrt{g}}\frac{\partial}{\partial v}\right)a(r,v)\right|_{r=r_{0}},
\end{equation}
or Dirichlet boundary conditions
\begin{equation}
0=\left.a(r,v)\right|_{r=r_{0}}.
\end{equation}
Note that the function $g=g(r,v)$ that appears here should not be confused with the metric determinant $g=\det(g_{\mu\nu})$.

\section{Amplitude expansion}

Assuming the source of the scalar field to be small enough ($\epsilon\ll1$), we can expand the fields in the amplitude of the disturbance:
\begin{align}
f(r,v)&=r+\sum_{n=1}^{\infty}\epsilon^{n}f_{n}(r,v), \\
g(r,v)&=r^{2}+\sum_{n=1}^{\infty}\epsilon^{n}g_{n}(r,v)
\end{align}
and
\begin{equation}
a(r,v)=\sum_{n=1}^{\infty}\epsilon^{n}a_{n}(r,v).
\end{equation}
For all $n$ we have the initial conditions
\begin{equation}
f_{n}(r,v)=0
\quad\text{,}\quad
g_{n}(r,v)=0
\quad\text{and}\quad
a_{n}(r,v)=0
\qquad\text{for}\qquad
v<0
\end{equation}
and the boundary conditions at the UV boundary ($r\rightarrow\infty$)
\begin{equation}
g_{n}(r,v)\leqslant\mathcal{O}\left(1\right)
\quad\text{,}\quad
f_{n}(r,v)\leqslant\mathcal{O}\left(\frac{1}{r}\right)
\quad\text{and}\quad
\begin{dcases}
a_{n}(r,v)\leqslant\mathcal{O}\left(\frac{1}{r}\right) & \text{ for } n>1, \\
\lim_{r\rightarrow\infty}a_{1}(r,v)=a_{0}(v).
\end{dcases}
\end{equation}
We can now solve the equations of motion order by order in $\epsilon$. This results in linear differential equations at each order. At zeroth order the equations of motion are solved by the background. At first order the equations of motion result in
\begin{equation}
-2(d-1)\partial_{r}^{2}f_{1}=0,
\end{equation}
\begin{equation}
\partial_{r}\left((d-2)f_{0}^{d-3}f_{1}\left(g_{0}\partial_{r}f_{0}+2\partial_{v}f_{0}\right)+f_{0}^{d-2}\left(g_{0}\partial_{r}f_{1}+g_{1}\partial_{r}f_{0}+2\partial_{v}f_{1}\right)\right)=d(d-1)f_{0}^{d-2}f_{1}
\end{equation}
and
\begin{equation}\label{eqn:FirstOrderEquationScalarField}
\partial_{r}\left(f_{0}^{d-1}g_{0}\partial_{r}a_{1}\right)+\partial_{r}\left(f_{0}^{d-1}\partial_{v}a_{1}\right)+\partial_{v}\left(f_{0}^{d-1}\partial_{r}a_{1}\right)=0.
\end{equation}
The first two of these can be easily solved by $f_{1}(r,v)=g_{1}(r,v)=0$, which implies that there is no backreaction on the metric at first order. So only the equation (\ref{eqn:FirstOrderEquationScalarField}) remains. Given this solution at first order the equations of motion at second order result in
\begin{equation}
-2(d-1)\partial_{r}^{2}f_{2}=f_{0}(\partial_{r}a_{1})^{2},
\end{equation}
\begin{equation}
\partial_{r}\left((d-2)f_{0}^{d-3}f_{2}\left(g_{0}\partial_{r}f_{0}+2\partial_{v}f_{0}\right)+f_{0}^{d-2}\left(g_{0}\partial_{r}f_{2}+g_{2}\partial_{r}f_{0}+2\partial_{v}f_{2}\right)\right)=d(d-1)f_{0}^{d-2}f_{2}
\end{equation}
and
\begin{equation}
\partial_{r}\left(f_{0}^{d-1}g_{0}\partial_{r}a_{2}\right)+\partial_{r}\left(f_{0}^{d-1}\partial_{v}a_{2}\right)+\partial_{v}\left(f_{0}^{d-1}\partial_{r}a_{2}\right)=0.
\end{equation}
This last equation has the simple solution $a_{2}=0$.

\section{Black brane formation}

In the absence of a hard wall, the authors of \cite{BM} find that in the translationally invariant setup the $\epsilon^{2}$-correction to the metric results in a Vaidya type metric
\begin{equation}
ds^{2}=-r^{2}\left(1-\frac{M(v)}{r^{d}}\right)dv^{2}+2drdv+r^{2}d\vec{x}^{2},
\end{equation}
where $M(v)\sim\frac{\epsilon^{2}}{(\delta t)^{d}}$. However the naive $\epsilon$-expansion of the solutions is not a good perturbation series since consequent corrections grow larger at late times instead of smaller even though $\epsilon\ll1$.%
\footnote{
Analogous expansions for shock-wave collisions in AdS backgrounds have been performed in the literature \cite{Grumiller:2008va,Lin:2010cb} where the region of applicability is also restricted to early times, unless re-summations techniques are employed as in \cite{Albacete:2009ji}.
}
If the perturbation is performed around the AdS-Vaidya background rather than around an AdS background, the $\epsilon$-expansion of the solutions results in a well behaved perturbation series. This leads to the conclusion that for arbitrary injection times $\delta t$ there is always a black brane formed with an event horizon $r_{h}\sim\frac{\epsilon^{2/d}}{\delta t}$. For times that are short compared to the inverse temperature of the black brane that is formed, the $\epsilon$-expansion around an AdS background is still valid.

Now we insert the hard wall again at a radial distance $r=r_{0}$. In the case that $r_{0}\lesssim\frac{\epsilon^{2/d}}{\delta t}$ we can follow the analysis of \cite{BM} and we find an infalling solution for the scalar field and a solution for the metric that describes the formation of a black brane with event horizon $r_{h}\sim\frac{\epsilon^{2/d}}{\delta t}$. The presence of the hard wall does not influence the leading behaviour of the exterior solution since it is well within the event horizon.

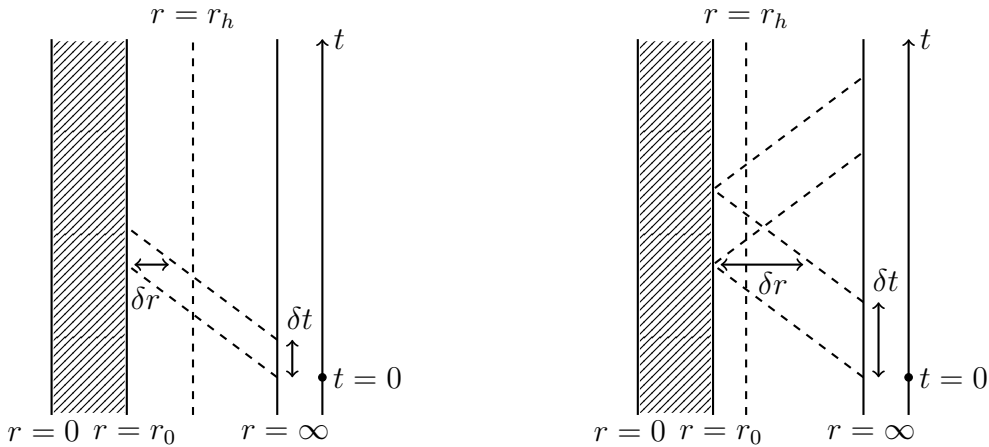
\begin{figure}[!h]
\centering
\begin{tikzpicture}
\draw[pattern=north east lines, pattern color=black] (0,0) rectangle (1,5);
\draw[ultra thick,color=white] (0,0) rectangle (1,5);
\draw[thick] (0,0) -- (0,5);
\draw (0,0.075) node [text=black,below] {$r=0\,\,\,$};
\draw[thick] (1,0) -- (1,5);
\draw (1,0) node [text=black,below] {$\,\,\, r=r_{0}$};
\draw[thick] (3,0) -- (3,5);
\draw (3,0) node [text=black,below] {$\,\,\, r=\infty$};
\draw[dashed,thick] (1.88,0) -- (1.88,5) node [text=black,above] {$r=r_{h}$};
\draw[<->,thick] (1.1,2) -- (1.56,2);
\draw (1.25,1.85) node [text=black,below] {$\delta r$};
\draw[dashed,thick] (3,0.5) -- (1,2);
\draw[dashed,thick] (3,1) -- (1,2.5);
\draw[<->,thick] (3.2,0.5) -- (3.2,1);
\draw (3.3,1) node [text=black,above] {$\delta t$};
\draw[->,thick] (3.6,0) -- (3.6,5) node [text=black,right] {$t$};
\draw[fill] (3.6,0.5) circle (0.05) node [text=black,right] {$t=0$};

\end{tikzpicture}
\hspace{2cm}
\begin{tikzpicture}
\draw[pattern=north east lines, pattern color=black] (0,0) rectangle (1,5);
\draw[ultra thick,color=white] (0,0) rectangle (1,5);
\draw[thick] (0,0) -- (0,5);
\draw (0,0.075) node [text=black,below] {$r=0\,\,\,$};
\draw[thick] (1,0) -- (1,5);
\draw (1,0) node [text=black,below] {$\,\,\, r=r_{0}$};
\draw[thick] (3,0) -- (3,5);
\draw (3,0) node [text=black,below] {$\,\,\, r=\infty$};
\draw[dashed,thick] (1.44,0) -- (1.44,5) node [text=black,above] {$r=r_{h}$};
\draw[<->,thick] (1.1,2) -- (2.2,2);
\draw (1.8,2.05) node [text=black,below] {$\delta r$};
\draw[dashed,thick] (3,0.5) -- (1,2) -- (3,3.5);
\draw[dashed,thick] (3,1.5) -- (1,3) -- (3,4.5);
\draw[<->,thick] (3.2,0.5) -- (3.2,1.5);
\draw (3.3,1.5) node [text=black,above] {$\delta t$};
\draw[->,thick] (3.6,0) -- (3.6,5) node [text=black,right] {$t$};
\draw[fill] (3.6,0.5) circle (0.05) node [text=black,right] {$t=0$};
\end{tikzpicture}
\caption{\emph{Explanation of the subleading effect due to the presence of the hard wall.
}}\label{fig:subleading}
\end{figure}

The main subleading effect due to the hard wall can be understood in the following way. Given an injection time $\delta t$, the spatial extent at a fixed time $t$ of the infalling shock wave near the horizon is $\delta r\sim r_{h}^{2}\delta t$. If the entire shell fits in the space between the hard wall and the horizon, thus as $r_{h}-r_{0}\gtrsim\delta r$ (see figure \ref{fig:subleading}), a black brane will be formed. However if $r_{h}-r_{0}\lesssim\delta r$, then a part of the shock wave will already have reflected on the hard wall out of the would-be black brane. Therefore one expects some scattering solution instead. Since the condition $r_{h}-r_{0}\gtrsim\delta r$ is equivalent to $r_{h}-r_{h}^{2}\delta t\gtrsim r_{0}$ or $\frac{\epsilon^{2/d}}{\delta t}(1-\epsilon^{2/d})\gtrsim r_{0}$, this is only a subleading effect.

\section{Scattering solution}

If the condition $r_{0}\lesssim\frac{\epsilon^{2/d}}{\delta t}$ is not satisfied, then we can not assume that the hard wall is always within an event horizon. Therefore we will have to incorporate its effect on the solutions.

\subsection{Scalar field solution}

The equation of motion for the first order correction to the scalar field, $a_{1}(r,v)$, corresponding to the probe limit, is given in (\ref{eqn:FirstOrderEquationScalarField}). With the background solution plugged in, this becomes
\begin{equation} \label{deax}
\partial_{r}\left(r^{d+1}\partial_{r}a_{1}\right)+\partial_{r}\left(r^{d-1}\partial_{v}a_{1}\right)+\partial_{v}\left(r^{d-1}\partial_{r}a_{1}\right)=0,
\end{equation}
If $d=2n+1$ is odd, then the two independent solutions to this equation can be written as the following finite sums.\footnote{More information on the scalar field solution can be found in Appendix \ref{ScalarFieldSolution}.} There is an
infalling solution (similar to the infalling solution found in \cite{BM})
\begin{equation}\label{aind}
a_{1}^{\text{in}}(r,v)=\sum_{k=0}^{n}\frac{2^{k}}{k!}\frac{\binom{n}{k}}{\binom{2n}{k}}\frac{1}{r^{k}}A^{(k)}(v)
\end{equation}
and an outgoing solution
\begin{equation}\label{aoutd}
a_{1}^{\text{out}}(r,v)=\sum_{k=0}^{n}\frac{2^{k}}{k!}\frac{\binom{n}{k}}{\binom{2n}{k}}\frac{(-1)^{k}}{r^{k}}B^{(k)}\left(v+\frac{2}{r}\right),
\end{equation}
for arbitrary functions $A(v)$ and $B(v)$. The names ``infalling'' and ``outgoing'' come from the fact that $v=t-\frac{1}{r}$ and $v+\frac{2}{r}=t+\frac{1}{r}$ are respectively the infalling and outgoing Eddington-Finkelstein coordinates. A systematic way to obtain the two solutions is the Frobenius method. We restrict the presentation to $d=3$ for concreteness. Then, one searches for solutions of the form
\begin{equation}\label{Frob}
a_{1}(r,v)=\sum_{k=0}^{\infty}\frac{C_{k}(v)}{r^{k}},
\end{equation}
from which the following recursion relation is obtained:
\begin{equation}\label{rec}
C_{k+1}(k+1)(k-2)=2(k-1)\partial_{v}C_{k}, \qquad k\geq0.
\end{equation}
The recursive relation forces $C_{2}=0$. Choosing $C_{3}=0$ and $C_{0}\neq0$ yields $C_{k}=0$ for $k\geq 2$ and the solution becomes
\begin{equation}\label{ain}
a_{1}^{\text{in}}(r,v)=A(v)+\frac{\dot{A}(v)}{r},
\end{equation}
where we renamed $C_{0}$ to $A$. Equation (\ref{ain}) is precisely the same as equation (\ref{aind}) for $d=3$. A second solution can be obtained by taking $C_{3}\neq0$ and $C_{0}=0$. Then, the $n^{\text{th}}$ term $C_{n}$ expressed in terms of $C_{3}$ yields
\begin{equation}
C_{n}=\frac{3(n-2)2^{n-2}}{n!}\partial_{v}^{n-3}C_{3}=\frac{(n-2)2^{n-1}}{n!}\partial_{v}^{n}\tilde{C}_{3}, \qquad n\geqslant3,
\end{equation}
where $3C_{3}/2=\partial_{v}^{3}\tilde{C}_{3}$ has been conveniently introduced. Dropping the tilde symbol from $\tilde{C}_{3}$ the second solution is given by
\begin{equation}\label{aout2}
a_{1}(r,v)=\frac{1}{2}\sum_{n=3}^{\infty}\left(\frac{2}{r}\right)^{n}\frac{(n-2)}{n!}\partial_{v}^{n}C_{3}=C_{3}+\frac{\dot{C}_{3}}{r}+   \left(\frac{\partial_v}{r}-1\right)e^{\frac{2}{r}\partial_{v}}C_{3}.
\end{equation}
The first two terms are in fact the same as the one independent solution (\ref{ain}). Observing that the exponential is a shift operator for the variable $v$ by $2/r$, the second independent solution is then
\begin{equation}\label{aout}
a_{1}^{\text{out}}(r,v)=B\left(v+\frac{2}{r}\right)-\frac{\dot{B}\left(v+\frac{2}{r}\right)}{r},
\end{equation}
which is the same as equation (\ref{aoutd}) for $d=3$. The general solution is then given by
\begin{equation}\label{aout}
a_{1}(r,v)=a_{1}^{\text{in}}(r,v)+a_{1}^{\text{out}}(r,v)=A(v)+\frac{\dot{A}(v)}{r}+B\left(v+\frac{2}{r}\right)-\frac{\dot{B}\left(v+\frac{2}{r}\right)}{r},
\end{equation}
for arbitrary functions $A(v)$ and $B(v)$. In what follows, suitable boundary conditions will be imposed at the hard wall (and at the UV boundary).

\subsubsection{Neumann boundary conditions}

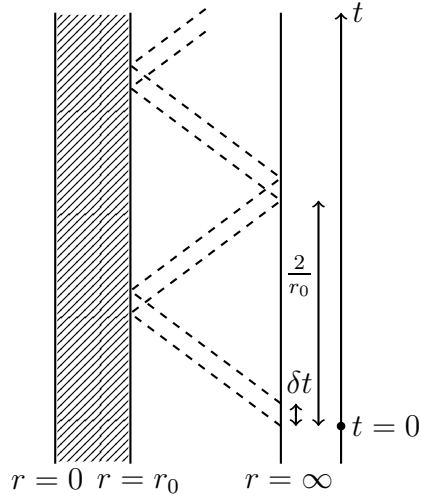
\begin{figure}[!h]
\centering
\begin{tikzpicture}
\draw[pattern=north east lines, pattern color=black] (0,0) rectangle (1,6);
\draw[ultra thick,color=white] (0,0) rectangle (1,6);
\draw[thick] (0,0) -- (0,6);
\draw (0,0.075) node [text=black,below] {$r=0\,\,\,$};
\draw[thick] (1,0) -- (1,6);
\draw (1,0) node [text=black,below] {$\,\,\, r=r_{0}$};
\draw[thick] (3,0) -- (3,6);
\draw (3,0) node [text=black,below] {$\,\,\, r=\infty$};
\draw[dashed,thick] (3,0.5) -- (1,2) -- (3,3.5) -- (1,5) -- (2,5.75);
\draw[dashed,thick] (3,0.8) -- (1,2.3) -- (3,3.8) -- (1,5.3) -- (2,6.05);
\draw[<->,thick] (3.2,0.5) -- (3.2,0.8);
\draw (3.25,0.8) node [text=black,above] {$\delta t$};
\draw[<->,thick] (3.5,0.5) -- (3.5,3.5);
\draw (3.575,2.5) node [text=black,left] {$\frac{2}{r_{0}}$};
\draw[->,thick] (3.8,0) -- (3.8,6) node [text=black,right] {$t$};
\draw[fill] (3.8,0.5) circle (0.05) node [text=black,right] {$t=0$};
\end{tikzpicture}
\caption{\emph{Schematic representation of the scattering solution in the bulk.
}}\label{fig:scattering}
\end{figure}

In particular for $d=3$, the UV boundary conditions require that
\begin{equation}
a_{0}(v)=\lim_{r\rightarrow\infty}a_{1}(r,v)=A(v)+B(v).
\end{equation}
If we impose Neumann boundary conditions on the hard wall, we find\footnote{The general Neumann boundary condition at the hard wall is (\ref{eqn:NeumannBoundaryCondition}), but since there is no backreaction to the metric at first order in $\epsilon$, we can safely take $g(r,v)=g_{0}(r,v)=r^{2}$ here.}
\begin{equation}
0=\left.\left(r\frac{\partial}{\partial r}+\frac{1}{r}\frac{\partial}{\partial v}\right)a_{1}(r,v)\right|_{r=r_{0}}=\frac{\ddot{A}(v)}{r_{0}^{2}}+\frac{\ddot{B}\left(v+\frac{2}{r_{0}}\right)}{r_{0}^{2}}.
\end{equation}
This implies that $A(v)+B\left(v+\frac{2}{r_{0}}\right)=C_{0}+C_{1}v$. The initial condition $a_{1}(r,v)=0$ for $v<0$ will determine $C_{0}=C_{1}=0$. Since the exponential of a derivative is a shift operator, we find the conditions
\begin{equation}
\begin{cases}
A(v)+B(v)=a_{0}(v), \\
A(v)+\exp\left(\frac{2}{r_{0}}\frac{\partial}{\partial v}\right)B(v)=0,
\end{cases}
\end{equation}
which can be solved by
\begin{align}
A(v)&=\left(\frac{\exp\left(\frac{2}{r_{0}}\frac{\partial}{\partial v}\right)}{\exp\left(\frac{2}{r_{0}}\frac{\partial}{\partial v}\right)-1}\right)a_{0}(v)=\left(\frac{1}{1-\exp\left(-\frac{2}{r_{0}}\frac{\partial}{\partial v}\right)}\right)a_{0}(v) \nonumber \\
&=\left(\sum_{n=0}^{\infty}\exp\left(-\frac{2n}{r_{0}}\frac{\partial}{\partial v}\right)\right)a_{0}(v)=\sum_{n=0}^{\infty}a_{0}\left(v-\frac{2n}{r_{0}}\right)
\end{align}
and
\begin{equation}
B(v)=-\exp\left(-\frac{2}{r_{0}}\frac{\partial}{\partial v}\right)A(v)=-A\left(v-\frac{2}{r_{0}}\right)=-\sum_{n=0}^{\infty}a_{0}\left(v-\frac{2(n+1)}{r_{0}}\right),
\end{equation}
such that
\begin{align}\label{eqn:ScalarSolutionNeumann}
a_{1}(r,v)=&\sum_{m=1}^{\infty}\left(a_{0}\left({\textstyle v-\frac{2(m-1)}{r_{0}}}\right)+\frac{\dot{a}_{0}\left(v-\frac{2(m-1)}{r_{0}}\right)}{r}\right. \nonumber \\
&\qquad\qquad\qquad\qquad\left.-a_{0}\left({\textstyle v+\frac{2}{r}-\frac{2m}{r_{0}}}\right)+\frac{\dot{a}_{0}\left(v+\frac{2}{r}-\frac{2m}{r_{0}}\right)}{r}\right).
\end{align}
This solution represents a scattering of infalling waves at the hard wall and the UV boundary. It is schematically depicted in figure \ref{fig:scattering}.

As a consistency check, it is easy to verify that in the limit $r_0\to 0$ in which the hard wall is removed, the scalar field solution of \cite{BM} for planar AdS$_4$ is recovered.

\subsubsection{Dirichlet boundary conditions}
\label{DirichletBoundaryConditions}

If instead we impose Dirichlet boundary conditions at the hard wall, we have
\begin{equation}
0=\left.a_{1}(r,v)\right|_{r=r_{0}}=A(v)+\frac{\dot{A}(v)}{r_{0}}+B\left(v+\frac{2}{r_{0}}\right)-\frac{\dot{B}\left(v+\frac{2}{r_{0}}\right)}{r_{0}}.
\end{equation}
Together with the UV boundary condition from last paragraph, we find the conditions\footnote{Equivalently, these can be solved by performing a Fourier transform $A(v)=\int_{-\infty}^{+\infty}\frac{\text{d}\omega}{2\pi}\,e^{i\omega v}A(\omega)$ such that the boundary conditions translate to
\begin{equation}
\begin{cases}
A(\omega)+B(\omega)=a_{0}(\omega), \\
\left(1+\frac{i\omega}{r_{0}}\right)A(\omega)+\exp\left(\frac{i2\omega}{r_{0}}\right)\left(1-\frac{i\omega}{r_{0}}\right)B(\omega)=0.
\end{cases}
\end{equation}}
\begin{equation}
\begin{cases}
A(v)+B(v)=a_{0}(v), \\
\left(1+\frac{1}{r_{0}}\frac{\partial}{\partial v}\right)A(v)+\exp\left(\frac{2}{r_{0}}\frac{\partial}{\partial v}\right)\left(1-\frac{1}{r_{0}}\frac{\partial}{\partial v}\right)B(v)=0.
\end{cases}
\end{equation}
These can be formally solved by
\begin{align}
A(v)&=\left(\frac{\exp\left(\frac{2}{r_{0}}\frac{\partial}{\partial v}\right)\left(1-\frac{1}{r_{0}}\frac{\partial}{\partial v}\right)}{\exp\left(\frac{2}{r_{0}}\frac{\partial}{\partial v}\right)\left(1-\frac{1}{r_{0}}\frac{\partial}{\partial v}\right)-\left(1+\frac{1}{r_{0}}\frac{\partial}{\partial v}\right)}\right)a_{0}(v) \nonumber \\
&=\left(1+\sum_{n=1}^{\infty}\exp\left(-\frac{2n}{r_{0}}\frac{\partial}{\partial v}\right)\left(1-\frac{1}{r_{0}}\frac{\partial}{\partial v}\right)^{-n}\left(1+\frac{1}{r_{0}}\frac{\partial}{\partial v}\right)^{n}\right)a_{0}(v).
\end{align}
In Appendix \ref{CalculationDirichletBoundaryCondition} it is shown that we can write
\begin{equation}
\left(1-\frac{1}{r_{0}}\frac{\partial}{\partial v}\right)^{-n}\left(1+\frac{1}{r_{0}}\frac{\partial}{\partial v}\right)^{n}f(v)=(-1)^{n}f(v)+\int_{0}^{\infty}\text{d}t\,F_{n}(t)f\left(v+\frac{t}{r_{0}}\right),
\end{equation}
where we have defined the functions $F_{0}(t)=0$ and $F_{n}(t)=\frac{1}{\Gamma(n)}\left(1-\frac{\partial}{\partial t}\right)^{n}\left(t^{n-1}e^{-t}\right)$ for $n\geqslant1$. The solution is therefore given by
\begin{equation}
A(v)=a_{0}(v)-B(v)=\sum_{n=0}^{\infty}\left[(-1)^{n}a_{0}\left(v-\frac{2n}{r_{0}}\right)+\int_{0}^{\infty}\text{d}t\,F_{n}(t)a_{0}\left(v+\frac{(t-2n)}{r_{0}}\right)\right].
\end{equation}
This also looks like a scattering solution where successive reflections are smeared out, as can be seen from the convolution integral in the solution. Following the calculations done in Appendix \ref{CalculationDirichletBoundaryCondition}, one can see that the smearing functions satisfy
\begin{equation}\label{eqn:SmearingFunction}
\int_{0}^{\infty}\text{d}t\,F_{n}(t)=1-(-1)^{n}.
\end{equation}

\subsection{Metric solution}

The equations of motion for the second order correction to the metric components $f_{2}(r,v)$ and $g_{2}(r,v)$ are given by
\begin{equation}
-2(d-1)\partial_{r}^{2}f_{2}=r(\partial_{r}a_{1})^{2}
\end{equation}
and
\begin{equation}
\partial_{r}\left((d-2)r^{d-1}f_{2}+r^{d}\partial_{r}f_{2}+r^{d-2}g_{2}+2r^{d-2}\partial_{v}f_{2}\right)=d(d-1)r^{d-2}f_{2}.
\end{equation}
These determine the leading backreaction of the metric to the scalar field. The general solution satisfying the initial and boundary conditions is given by
\begin{equation}\label{eqn:MetricFsolution}
f_{2}(r,v)=\frac{1}{2(d-1)}\left(r\int_{r}^{\infty}\rho(\partial_{\rho}a_{1}(\rho,v))^{2}\text{d}\rho-\int_{r}^{\infty}\rho^{2}(\partial_{\rho}a_{1}(\rho,v))^{2}\text{d}\rho\right)
\end{equation}
and
\begin{align}\label{eqn:MetricGsolution}
g_{2}(r,v)=&-2\partial_{v}f_{2}(r,v)-\frac{r}{(d-1)}\int_{r}^{\infty}\rho^{2}(\partial_{\rho}a_{1}(\rho,v))^{2}\text{d}\rho \nonumber \\
&-\frac{1}{2(d-1)r^{d-2}}\int_{r_{0}}^{r}\rho^{d+1}(\partial_{\rho}a_{1}(\rho,v))^{2}\text{d}\rho+\frac{\beta(v)}{r^{d-2}}.
\end{align}
By demanding that $E_{vv}=0$ at $r=r_{0}$, we find that the function $\beta(v)$ is fixed to be
\begin{equation}
\beta(v)=-\frac{r_{0}^{d-1}}{(d-1)}\int_{0}^{v}\left((\partial_{w}a_{1}(r_{0},w))^{2}+r_{0}^{2}(\partial_{r}a_{1}(r_{0},w))(\partial_{w}a_{1}(r_{0},w))\right)\text{d}w.
\end{equation}

\subsection{Regime of validity of perturbation theory}
\label{Regime of validity of perturbation theory}

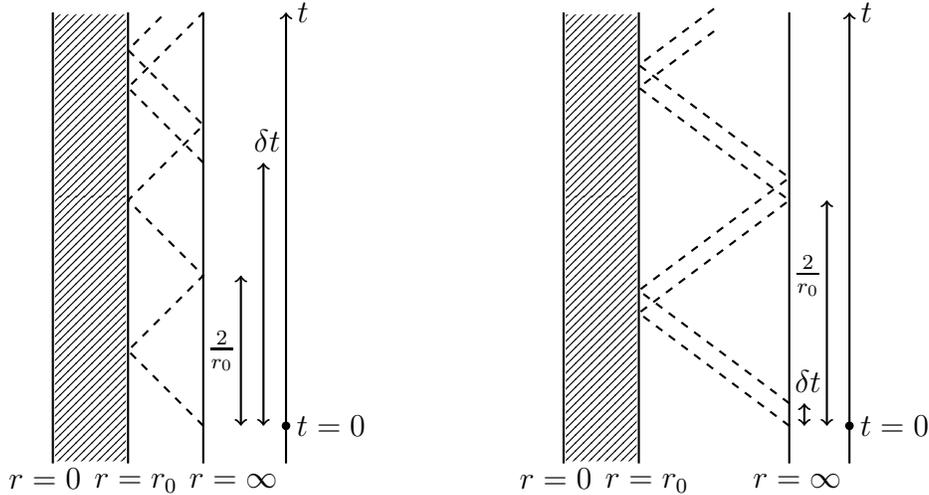
\begin{figure}[!h]
\centering
\begin{tikzpicture}
\draw[pattern=north east lines, pattern color=black] (0,0) rectangle (1,6);
\draw[ultra thick,color=white] (0,0) rectangle (1,6);
\draw[thick] (0,0) -- (0,6);
\draw (0,0.075) node [text=black,below] {$r=0\,\,\,$};
\draw[thick] (1,0) -- (1,6);
\draw (1,0) node [text=black,below] {$\,\,\,r=r_{0}$};
\draw[thick] (2,0) -- (2,6);
\draw (2.4,0) node [text=black,below] {$r=\infty$};
\draw[dashed,thick] (2,0.5) -- (1,1.5) -- (2,2.5) -- (1,3.5) -- (2,4.5) -- (1,5.5) -- (1.5,6);
\draw[dashed,thick] (2,4) -- (1,5) -- (2,6);
\draw[<->,thick] (2.8,0.5) -- (2.8,4);
\draw (2.85,4) node [text=black,above] {$\delta t$};
\draw[<->,thick] (2.5,0.5) -- (2.5,2.5);
\draw (2.575,1.5) node [text=black,left] {$\frac{2}{r_{0}}$};
\draw[->,thick] (3.1,0) -- (3.1,6) node [text=black,right] {$t$};
\draw[fill] (3.1,0.5) circle (0.05) node [text=black,right] {$t=0$};
\end{tikzpicture}
\hspace{1.5cm}
\begin{tikzpicture}
\draw[pattern=north east lines, pattern color=black] (0,0) rectangle (1,6);
\draw[ultra thick,color=white] (0,0) rectangle (1,6);
\draw[thick] (0,0) -- (0,6);
\draw (0,0.075) node [text=black,below] {$r=0\,\,\,$};
\draw[thick] (1,0) -- (1,6);
\draw (1,0) node [text=black,below] {$\,\,\, r=r_{0}$};
\draw[thick] (3,0) -- (3,6);
\draw (3,0) node [text=black,below] {$\,\,\, r=\infty$};
\draw[dashed,thick] (3,0.5) -- (1,2) -- (3,3.5) -- (1,5) -- (2,5.75);
\draw[dashed,thick] (3,0.8) -- (1,2.3) -- (3,3.8) -- (1,5.3) -- (2,6.05);
\draw[<->,thick] (3.2,0.5) -- (3.2,0.8);
\draw (3.25,0.8) node [text=black,above] {$\delta t$};
\draw[<->,thick] (3.5,0.5) -- (3.5,3.5);
\draw (3.575,2.5) node [text=black,left] {$\frac{2}{r_{0}}$};
\draw[->,thick] (3.8,0) -- (3.8,6) node [text=black,right] {$t$};
\draw[fill] (3.8,0.5) circle (0.05) node [text=black,right] {$t=0$};
\end{tikzpicture}
\caption{\emph{(Left) Scattering with $r_{0}\delta t\gg 1$. (Right) Scattering with $r_{0}\delta t\ll 1$.
}}
\label{fig:twoscatteringsolutions}
\end{figure}

We will now investigate under what conditions the scattering solution that we found here results in a consistent perturbation series. This analysis is done for the solution with $d=3$ and Neumann boundary conditions. (For Dirichlet boundary conditions, the analysis can be done similarly.) The main condition that needs to be satisfied is that the corrections to the metric coefficients should be small compared to their background values, i.e. $\epsilon^{2}f_{2}(r,v)\ll f_{0}(r,v)$ and $\epsilon^{2}g_{2}(r,v)\ll g_{0}(r,v)$. We start by separating the terms $a_{1}(r,v)=a_{1,1}(r,v)+a_{1,2}(r,v)$ in solution (\ref{eqn:ScalarSolutionNeumann}) as
\begin{equation}
\epsilon\,a_{1,1}(r,v)=\epsilon\sum_{m=1}^{\infty}\left(a_{0}\left({\textstyle v-\frac{2(m-1)}{r_{0}}}\right)-a_{0}\left({\textstyle v+\frac{2}{r}-\frac{2m}{r_{0}}}\right)\right)
\end{equation}
and
\begin{equation}
\epsilon\,a_{1,2}(r,v)=\epsilon\sum_{m=1}^{\infty}\left(\frac{\dot{a}_{0}\left(v-\frac{2(m-1)}{r_{0}}\right)}{r}+\frac{\dot{a}_{0}\left(v+\frac{2}{r}-\frac{2m}{r_{0}}\right)}{r}\right).
\end{equation}
Furthermore, note that we can separate $\partial_{r}a_{1}(r,v)=b_{1}(r,v)+b_{2}(r,v)$ as
\begin{equation}
\epsilon\,b_{1}(r,v)=\epsilon\sum_{m=1}^{\infty}\left(\frac{\dot{a}_{0}\left(v+\frac{2}{r}-\frac{2m}{r_{0}}\right)}{r^{2}}-\frac{\dot{a}_{0}\left(v-\frac{2(m-1)}{r_{0}}\right)}{r^{2}}\right)
\end{equation}
and
\begin{equation}
\epsilon\,b_{2}(r,v)=-2\epsilon\sum_{m=1}^{\infty}\frac{\ddot{a}_{0}\left(v+\frac{2}{r}-\frac{2m}{r_{0}}\right)}{r^{3}}.
\end{equation}
Since the function $a_{0}(v)$ has compact support, the relevant summation is over terms that have $0<\frac{2\Delta m}{r_{0}}<\delta t$. If $r_{0}\delta t\gg 1$ the number of terms in the sum that we need to take into account is estimated by $N\sim r_{0}\delta t$. If $r_{0}\delta t\ll 1$, then we have only one relevant term and thus $N\sim 1$. Both possibilities are schematically represented in figure \ref{fig:twoscatteringsolutions}, and will now be discussed in more detail.

\subsubsection{Estimate for $r_{0}\delta t\gg 1$}

The minus sign in the expression of $a_{1,1}$ is crucial since it will cause a cancelation of different terms. If we assume that $a_{0}(v)$ is a sufficiently smooth function, then noting that $\frac{1}{r}\leqslant\frac{1}{r_{0}}\ll\delta t$, we find that in the expression of $a_{1,1}$ for every $m$ the term $a_{0}\left({\textstyle v-\frac{2(m-1)}{r_{0}}}\right)$ will (approximately) cancel the term $a_{0}\left({\textstyle v+\frac{2}{r}-\frac{2m}{r_{0}}}\right)$. We can thus estimate that
\begin{equation}
\epsilon\,a_{1,1}\sim\epsilon N-\epsilon N\lesssim\epsilon,
\end{equation}
is at most of order $\epsilon$. Since the function $f(r)=1/r$ is maximal at $r=r_{0}$ we also have that
\begin{equation}
\epsilon\,a_{1,2}\sim\epsilon\sum_{m}\frac{\dot{a}_{0}\left(v-\frac{2m}{r_{0}}\right)}{r_{0}}\sim\frac{\epsilon N}{r_{0}\delta t}\sim\epsilon.
\end{equation}
In a similar way we can estimate
\begin{equation}
\epsilon\,b_{1}(r,v)\sim\frac{\epsilon N}{r_{0}^{2}\delta t}-\frac{\epsilon N}{r_{0}^{2}\delta t}\lesssim\frac{\epsilon}{r_{0}^{2}\delta t}
\end{equation}
and
\begin{equation}
\epsilon\,b_{2}(r,v)\sim\epsilon\sum_{m}\frac{\ddot{a}_{0}\left(v-\frac{2m}{r_{0}}\right)}{r_{0}^{3}}\sim\frac{\epsilon N}{r_{0}^{3}\delta t^{2}}\sim\frac{\epsilon}{r_{0}^{2}\delta t},
\end{equation}
such that from expression (\ref{eqn:MetricFsolution}) we get
\begin{equation}
\epsilon^{2}\left.f_{2}\right|_{r=r_{0}}\sim r_{0}^{3}\left.(\epsilon\,\partial_{r}a)^{2}\right|_{r=r_{0}}\sim\frac{\epsilon^{2}}{r_{0}\delta t^{2}}.
\end{equation}
Because $f_{0}=r$, the condition $\epsilon^{2}f_{2}(r,v)\ll f_{0}(r,v)$ thus implies $\epsilon\ll r_{0}\delta t$. The solution for $g_{2}(r,v)$ contains three different kind of terms. The fourth term in (\ref{eqn:MetricGsolution}) can be estimated by
\begin{align}
\epsilon^{2}\left.g_{2}\right|_{r=r_{0}}&\gtrsim r_{0}\left.\left((\epsilon\,\partial_{v}a)^{2}+r_{0}^{2}(\epsilon\,\partial_{r}a)(\epsilon\,\partial_{v}a)\right)\delta t\right|_{r=r_{0}} \nonumber \\
&\sim r_{0}\delta t\left(\left(\frac{\epsilon}{\delta t}\right)^{2}+r_{0}^{2}\left(\frac{\epsilon}{r_{0}^{2}\delta t}\right)\left(\frac{\epsilon}{\delta t}\right)\right)\sim\frac{r_{0}\epsilon^{2}}{\delta t}.
\end{align}
The second and third term in (\ref{eqn:MetricGsolution}) can be estimated by
\begin{equation}
\epsilon^{2}\left.g_{2}\right|_{r=r_{0}}\gtrsim r_{0}^{4}\left.(\epsilon\,\partial_{r}a)^{2}\right|_{r=r_{0}}\sim\frac{\epsilon^{2}}{\delta t^{2}}.
\end{equation}
Finally, the first term in (\ref{eqn:MetricGsolution}) can be estimated by
\begin{equation}
\epsilon^{2}\left.g_{2}\right|_{r=r_{0}}\gtrsim\epsilon^{2}\left.\partial_{v}f_{2}\right|_{r=r_{0}}\sim\frac{\epsilon^{2}}{r_{0}\delta t^{3}}.
\end{equation}
Because $g_{0}=r^{2}$, the condition $\epsilon^{2}g_{2}(r,v)\ll g_{0}(r,v)$ thus implies $\epsilon\ll(r_{0}\delta t)^{\frac{1}{2}}$, $\epsilon\ll r_{0}\delta t$ and $\epsilon\ll(r_{0}\delta t)^{\frac{3}{2}}$. Since $r_{0}\delta t\gg1$ it is sufficient to have the condition $\epsilon\ll(r_{0}\delta t)^{\frac{1}{2}}$. However, as $\epsilon\ll 1$ and $r_{0}\delta t\gg 1$, this condition is automatically satisfied and our scattering solution is always valid. This could have been anticipated from the fact that very slow injection times ($\delta t\gg1/r_{0}$) lead to adiabatic changes in the bulk.

\subsubsection{Estimate for $r_{0}\delta t\ll 1$}

Since in this case $N\sim1$, we find that
\begin{equation}
\epsilon\,a_{1,1}\sim\epsilon N-\epsilon N\lesssim\epsilon,
\end{equation}
and since the function $f(r)=1/r$ is maximal at $r=r_{0}$ also
\begin{equation}
\epsilon\,a_{1,2}\sim\epsilon\sum_{m}\frac{\dot{a}_{0}\left(v-\frac{2m}{r_{0}}\right)}{r_{0}}\sim\frac{\epsilon N}{r_{0}\delta t}\sim\frac{\epsilon}{r_{0}\delta t}.
\end{equation}
Note that since $r_{0}\delta t\ll 1$, we have that $\epsilon\,a_{1,1}\ll\epsilon\,a_{1,2}$. In a similar way we can estimate
\begin{equation}
\epsilon\,b_{1}(r,v)\sim\frac{\epsilon N}{r_{0}^{2}\delta t}-\frac{\epsilon N}{r_{0}^{2}\delta t}\lesssim\frac{\epsilon}{r_{0}^{2}\delta t}
\end{equation}
and
\begin{equation}
\epsilon\,b_{2}(r,v)\sim\epsilon\sum_{m}\frac{\ddot{a}_{0}\left(v-\frac{2m}{r_{0}}\right)}{r_{0}^{3}}\sim\frac{\epsilon N}{r_{0}^{3}\delta t^{2}}\sim\frac{\epsilon}{r_{0}^{3}\delta t^{2}},
\end{equation}
such that $\epsilon\,b_{1}\ll\epsilon\,b_{2}$ and from expression (\ref{eqn:MetricFsolution}) we get
\begin{equation}
\epsilon^{2}\left.f_{2}\right|_{r=r_{0}}\sim r_{0}^{3}\left.(\epsilon\,\partial_{r}a)^{2}\right|_{r=r_{0}}\sim\frac{\epsilon^{2}}{r_{0}^{3}\delta t^{4}}.
\end{equation}
Because $f_{0}=r$, the condition $\epsilon^{2}f_{2}(r,v)\ll f_{0}(r,v)$ thus implies $\epsilon\ll(r_{0}\delta t)^{2}$. The solution for $g_{2}(r,v)$ contains three different kind of terms. The fourth term in (\ref{eqn:MetricGsolution}) can be estimated by
\begin{align}
\epsilon^{2}\left.g_{2}\right|_{r=r_{0}}&\gtrsim r_{0}\left.\left((\epsilon\,\partial_{v}a)^{2}+r_{0}^{2}(\epsilon\,\partial_{r}a)(\epsilon\,\partial_{v}a)\right)\delta t\right|_{r=r_{0}} \nonumber \\
&\sim r_{0}\delta t\left(\left(\frac{\epsilon}{r_{0}\delta t^{2}}\right)^{2}+r_{0}^{2}\left(\frac{\epsilon}{r_{0}^{3}\delta t^{2}}\right)\left(\frac{\epsilon}{r_{0}\delta t^{2}}\right)\right)\sim\frac{\epsilon^{2}}{r_{0}\delta t^{3}}.
\end{align}
The second and third term in (\ref{eqn:MetricGsolution}) can be estimated by
\begin{equation}
\epsilon^{2}\left.g_{2}\right|_{r=r_{0}}\gtrsim r_{0}^{4}\left.(\epsilon\,\partial_{r}a)^{2}\right|_{r=r_{0}}\sim\frac{\epsilon^{2}}{r_{0}^{2}\delta t^{4}}.
\end{equation}
Finally, the first term in (\ref{eqn:MetricGsolution}) can be estimated by
\begin{equation}
\epsilon^{2}\left.g_{2}\right|_{r=r_{0}}\gtrsim\epsilon^{2}\left.\partial_{v}f_{2}\right|_{r=r_{0}}\sim\frac{\epsilon^{2}}{r_{0}^{3}\delta t^{5}}.
\end{equation}
Because $g_{0}=r^{2}$, the condition $\epsilon^{2}g_{2}(r,v)\ll g_{0}(r,v)$ thus implies that $\epsilon\ll(r_{0}\delta t)^{\frac{3}{2}}$, $\epsilon\ll(r_{0}\delta t)^{2}$ and $\epsilon\ll(r_{0}\delta t)^{\frac{5}{2}}$. Since $r_{0}\delta t\ll1$ it is sufficient to have the condition $\epsilon\ll(r_{0}\delta t)^{\frac{5}{2}}$. We can compare this with the condition $\epsilon^{2}\gtrsim(r_{0}\delta t)^{3}$ for having black brane formation in the bulk and note that these requirements are compatible with two well-separated regimes of validity. This result was schematically pictured in figure \ref{fig:TwoRegimes}.

\addcontentsline{toc}{section}{Acknowledgments}
\acknowledgments

We would like to thank W.~van der Schee  for a helpful discussion.

This work was supported in part by the Belgian Federal Science Policy Office through the Interuniversity Attraction Poles P7/37, by FWO-Vlaanderen through projects G011410N and G020714N,  by the Vrije Universiteit Brussel through the Strategic Research Program ``High-Energy Physics'', by European Union's Seventh Framework Programme under grant agreements (FP7-REGPOT-2012-2013-1) no 316165, PIF-GA-2011-300984, the ERC Advanced Grant BSMOXFORD 228169, the EU program ``Thales'' MIS 375734  and was also co-financed by the European Union (European Social Fund, ESF) and Greek national funds through the Operational Program ``Education and Lifelong Learning'' of the National Strategic Reference Framework (NSRF) under ``Funding of proposals that have received a positive evaluation in the 3rd and 4th Call of ERC Grant Schemes''. The work of AT is funded by the VUB Research Council. JV is Aspirant FWO. BC and JV acknowledge support from the Erasmus Intensive Programme ``Non-Perturbative Quantum Field Theory'' at the CCTP, where this project was initiated. HZ would like to acknowledge the wonderful hospitality of the CCTP in the final stages of this project.

\newpage
 \addcontentsline{toc}{section}{Appendices}
  \renewcommand{\theequation}{\thesection.\arabic{equation}}
\appendix
\section{Review of weak-field black hole formation in global AdS}
\label{reviewBM}

In this appendix, we briefly review the analysis of \cite{BM} for the case of global AdS. The bulk metric and scalar are written in the form
\bea
ds^2&=&2dr\,dv-g(r,v)\,dv^2+f^2(r,v)\,d\Omega^2_{d-1};  \\
a&=&a(r,v).
\eea
We impose pure AdS initial conditions
\bea
g(r,v)&=&r^2+\frac{1}{R^2}\ \ \ (v<0);\\
f(r,v)&=&rR\ \ \ (v<0);\\
a(r,v)&=&0\ \ \ (v<0),
\eea
and large $r$ boundary conditions
\bea
g(r,v)&=&r^2\left(1+{\cal O}\left(\frac{1}{r^2}\right)\right);\\
f(r,v)&=&r\left(R+{\cal O}\left(\frac{1}{r^2}\right)\right);\\
a(r,v)&=&a_0(v)+{\cal O}\left(\frac{1}{r}\right).
\eea
The independent equations of motion are the dynamical scalar field equation%
\footnote{
Note that there appears to be a typo in (4.4) of \cite{BM}, namely an extra factor of $g$ in the last term of the LHS.
}
\be
\partial_r(f^{d-1}g\partial_r a)+\partial_v(f^{d-1}\partial_r a)+\partial_r(f^{d-1}\partial_v a)=0
\ee
and two constraint equations determining the metric coefficients if the scalar field is known,
\bea
&&(\partial_r a)^2+\frac{2(d-1)\partial_r^2f}{f}=0;             \\
&&\partial_r(f^{d-2}g\partial_rf+2f^{d-2}\partial_vf)-d\,f^{d-1}-(d-2)f^{d-3}=0.
\eea
These equations have to be supplemented by an energy conservation equation at one value of $r$, which relates two functions that are undetermined by a large $r$ expansion of the equations of motion.

We wish to solve these equations in an amplitude expansion (which will effectively linearize the equations of motion)
\bea
a(r,v)&=&\sum_{n=0}^\infty \epsilon^n a_n(r,v);\\
f(r,v)&=&\sum_{n=0}^\infty \epsilon^nf_n(r,v);\\
g(r,v)&=&\sum_{n=0}^\infty \epsilon^ng_n(r,v),
\eea
with
\be
a_0(r,v)=0,\ \ \ f_0(r,v)=rR,\ \ \ g_0(r,v)=r^2+\frac{1}{R^2},
\ee
and the forcing function $a_0(v)$ is taken to be of order $\epsilon$.

We specialize to the case $d=3$ (global AdS$_4$) and consider the scalar field equation of motion at order $\epsilon$ (the metric equations are trivial at this order, since backreaction only occurs at order $\epsilon^2$):
\be\label{dilatoneq}
\partial_r\left[r^2R^2\left(r^2+\frac{1}{R^2}\right)\partial_r a_1\right]+\partial_v\left(r^2R^2\partial_r a_1\right)
+\partial_r\left(r^2R^2\partial_v a_1\right)=0.
\ee
To solve it, we expand the field $a_1$ in powers of $1/r$:
\be\label{seriesk}
a_1(r,v)=\sum_{k=0}^\infty\frac{a_{1,k}(v)}{r^k}.
\ee
The equation of motion then reduces to the recursion relation
\be
k(k-3)a_{1,k}-2(k-2)\dot{a}_{1,k-1}+\frac{(k-2)(k-3)}{R^2}a_{1,k-2}=0.
\ee
Given that $a_{1,0}=a_0(v)$, we find that
\be
a_{1,1}=\dot{a}_0(v), \ \ \ a_{1,2}=0,
\ee
while $a_{1,3}$ is undetermined. Given a choice for $a_{1,3}(v)$, the recursion relation determines the higher coefficients, e.g.,
\be
a_{1,4}=\dot{a}_{1,3}.
\ee
The only choice for which the series \eq{seriesk} truncates is $a_{1,3}(v)=0$, so that
\be\label{phi1}
a_1=a_0(v)+\frac{\dot{a}_0(v)}{r}.
\ee
This solution is manifestly infalling, and one can expect it to be the relevant solution in situations where the infalling shell forms a black hole (without scattering back towards the boundary at this order in the amplitude expansion). Indeed, in section 4.4 of \cite{BM}, it has been verified that this choice leads to a reliable perturbation expansion in a certain regime of parameters (namely the regime in which a black hole is formed). Specifically, in addition to the small parameter $\epsilon$, introduce the parameter
\be
x\equiv\frac{\delta t}{R},
\ee
where $\delta t$ is the duration of energy injection ($a_0(v)$ is only nonzero for $0<v<\delta t$) and $R$ is the radius of the $S^2$ of the boundary field theory. The result of \cite{BM}, where it is always assumed that $x\ll1$, is as follows:
\begin{itemize}
\item
If $x\ll\epsilon^{2/3}$, the horizon radius is
\be
r_H\sim\frac{\epsilon^{2/3}}{\delta t}\gg\frac1R.
\ee
A large black hole is formed and the naive perturbation theory described above is good as long as $vT\ll 1$, i.e., for times small compared to the inverse temperature of the black hole to be formed.
\item
If  $x\gg\epsilon^{2/3}$, the horizon radius of the metric following from \eq{phi1} is
\be
r_H\sim\frac{\epsilon^{2}}{x^3}\frac1R\ll\frac1R,
\ee
corresponding to a small black hole. Naive perturbation theory is good when $vT\ll 1$ for $v/R\ll\epsilon^2/x^3$, which is always obeyed for $v\sim\delta t$ as long as
\be
x\ll\sqrt\epsilon.
\ee
So in this regime we know that right after the injection of energy has ended, the bulk metric is well-approximated by a black hole geometry. In the opposite regime $x\gg\sqrt\epsilon$, the perturbative solution corresponding to \eq{phi1} is not valid. Indeed, we will see below that the actual solution in that regime is very different.
\end{itemize}

In section 4.2 of \cite{BM}, another solution to \eq{dilatoneq} is considered, which is regular everywhere and turns out to be a starting point for a good perturbation theory in the regime where no black hole is formed:
\bea
a_1(r,v)&=&\sum_{m=0}^\infty(-1)^m\left[a_0(v-m\pi R)+\frac{\dot{a}_0(v-m\pi R)}{r} +a_0(v-m\pi R-2R\arctan(rR))\right.\nonumber\\
&&\ \ \ \ \ \ \ \ \ \ \ \ \ \ \left. -\frac{\dot{a}_0(v-m\pi R-2R\arctan(rR))}{r}\right].\label{phi1regular}
\eea
To get some intuition for this expression, note that the metric in the $(r,v)$ plane can be written as
\be
-\left(r^2+\frac{1}{R^2}\right)\,dv\,d(v-2R\arctan(rR)),
\ee
so that the above expression is a combination of shells falling towards and scattering away from the center of AdS. The perturbation expansion based on \eq{phi1regular} is valid if $\phi_1$ is everywhere small, which is the case provided that
\be
\epsilon\ll1,\ \ \ x\gg\sqrt\epsilon.
\ee

\section{Scalar field solution}
\label{ScalarFieldSolution}

In this appendix, we analyze further the solution of equation (\ref{deax}) for the scalar field. In general the solution of the equation
\begin{equation}
\partial_{r}\left(r^{d+1}\partial_{r}a_{1}\right)+\partial_{r}\left(r^{d-1}\partial_{v}a_{1}\right)+\partial_{v}\left(r^{d-1}\partial_{r}a_{1}\right)=0
\end{equation}
can be found from a $1/r$-expansion
\begin{equation}\label{Eqn:overRexpansion}
a_{1}(r,v)=a_{0}(v)+\frac{\dot{a}_{0}(v)}{r}+...+\frac{L(v)}{r^{d}}+...\,,
\end{equation}
where the dot on $a_0$ denotes differentiation with respect to $v$. If $d=2n+1$ is odd, then all terms to the right of $\frac{L(v)}{r^{d}}$ contain only derivatives of $L(v)$ but not of $a_{0}(v)$. This expansion can thus be written as a finite sum involving only $a_{0}(v)$ plus an infinite series involving only $L(v)$. The terms can be summed explicitly in the following expression:
\begin{equation}
a_{1}(r,v)=\sum_{k=0}^{n}\frac{2^{k}}{k!}\frac{\binom{n}{k}}{\binom{2n}{k}}\frac{a_{0}^{(k)}(v)}{r^{k}}+\int_{-\infty}^{+\infty}\frac{P\left(v-\frac{t}{r}\right)\text{d}t}{(t(t+2))^{n+1}}
\quad\text{with}\quad
L(v)=\int_{-\infty}^{+\infty}\frac{P(v-w)}{w^{d+1}}\text{d}w.
\label{eq:solutionscalar}
\end{equation}
The agreement of the second term with the general form of the solution as presented in (\ref{Eqn:overRexpansion}) can be seen by expanding it in powers of $1/r$
\begin{align}
\int_{-\infty}^{+\infty}\frac{P\left(v-\frac{t}{r}\right)\text{d}t}{(t(t+2))^{n+1}}&=\int_{-\infty}^{+\infty}\frac{rP(v-w)\text{d}w}{(rw(rw+2))^{n+1}} \nonumber \\
&=\int_{-\infty}^{+\infty}\left[\frac{1}{(wr)^{d}}+\mathcal{O}\left(\frac{1}{(wr)^{d+1}}\right)\right]P(v-w)\frac{\text{d}w}{w}.
\end{align}
Note that solution (\ref{eq:solutionscalar}) corresponds to the boundary condition $\lim_{r\rightarrow\infty}a_{1}(r,v)=a_{0}(v)$ at the UV boundary.

\section{Calculation with Dirichlet boundary condition}
\label{CalculationDirichletBoundaryCondition}

In this appendix, we elaborate on the calculations that are needed in section \ref{DirichletBoundaryConditions}, when dealing with Dirichlet boundary conditions. For $n\geqslant1$, we can use the integral identity $\frac{1}{\Delta^{\alpha}}=\frac{1}{\Gamma(\alpha)}\int_{0}^{\infty}\text{d}t\,t^{\alpha-1}e^{-t\Delta}$ to write
\begin{align}
X&\equiv\left(1-\frac{1}{r_{0}}\frac{\partial}{\partial v}\right)^{-n}\left(1+\frac{1}{r_{0}}\frac{\partial}{\partial v}\right)^{n}f(v)=\left(1+\frac{1}{r_{0}}\frac{\partial}{\partial v}\right)^{n}\left(1-\frac{1}{r_{0}}\frac{\partial}{\partial v}\right)^{-n}f(v) \nonumber \\
&=\left(1+\frac{1}{r_{0}}\frac{\partial}{\partial v}\right)^{n}\frac{1}{\Gamma(n)}\int_{0}^{\infty}\text{d}t\,t^{n-1}\exp\left(-t\left(1-\frac{1}{r_{0}}\frac{\partial}{\partial v}\right)\right)f(v) \nonumber \\
&=\frac{1}{\Gamma(n)}\int_{0}^{\infty}\text{d}t\,t^{n-1}e^{-t}\left(1+\frac{1}{r_{0}}\frac{\partial}{\partial v}\right)^{n}\exp\left(\frac{t}{r_{0}}\frac{\partial}{\partial v}\right)f(v).
\end{align}
Remember that the exponential is a shift operator of the argument,
\begin{equation}
\exp\left(\frac{t}{r_{0}}\frac{\partial}{\partial v}\right)f(v)=\sum_{k=0}^{\infty}\frac{1}{k!}\left(\frac{t}{r_{0}}\frac{\partial}{\partial v}\right)^{k}f(v)=\sum_{k=0}^{\infty}\left(\frac{t}{r_{0}}\right)^{k}\frac{f^{(k)}(v)}{k!}=f\left(v+\frac{t}{r_{0}}\right).
\end{equation}
We can now proceed to compute
\begin{align}
X&=\frac{1}{\Gamma(n)}\int_{0}^{\infty}\text{d}t\,t^{n-1}e^{-t}\left(1+\frac{1}{r_{0}}\frac{\partial}{\partial v}\right)^{n}f\left(v+\frac{t}{r_{0}}\right) \nonumber \\
&=\frac{1}{\Gamma(n)}\int_{0}^{\infty}\text{d}t\,t^{n-1}e^{-t}\sum_{k=0}^{n}\binom{n}{k}\left(\frac{1}{r_{0}}\frac{\partial}{\partial v}\right)^{k}f\left(v+\frac{t}{r_{0}}\right) \nonumber \\
&=\frac{1}{\Gamma(n)}\sum_{k=0}^{n}\binom{n}{k}\int_{0}^{\infty}\text{d}t\,t^{n-1}e^{-t}\left(\frac{\partial}{\partial t}\right)^{k}f\left(v+\frac{t}{r_{0}}\right).
\end{align}
We note that $\partial_{t}^{k}(t^{n-1}e^{-t})=\left((n-k)...(n-2)(n-1)t^{n-k-1}+\mathcal{O}(t^{n-k})\right)e^{-t}$, such that $\lim_{t\rightarrow\infty}\partial_{t}^{k}(t^{n-1}e^{-t})=0$ and
\begin{equation}
\lim_{t\rightarrow0}\partial_{t}^{k}(t^{n-1}e^{-t})=\begin{cases}
0 & \text{ if } k<n-1, \\
\Gamma(n) & \text{ if } k=n-1.
\end{cases}
\end{equation}
Therefore, the result of partial integration is given by
\begin{align}
X&=(-1)^{n}f(v)+\frac{1}{\Gamma(n)}\sum_{k=0}^{n}\binom{n}{k}\int_{0}^{\infty}\text{d}t\left(-\frac{\partial}{\partial t}\right)^{k}\left(t^{n-1}e^{-t}\right)f\left(v+\frac{t}{r_{0}}\right) \nonumber \\
&=(-1)^{n}f(v)+\frac{1}{\Gamma(n)}\int_{0}^{\infty}\text{d}t\left(1-\frac{\partial}{\partial t}\right)^{n}\left(t^{n-1}e^{-t}\right)f\left(v+\frac{t}{r_{0}}\right).
\end{align}
This is the result that was stated in section \ref{DirichletBoundaryConditions}. Finally, we note that
\begin{align}
&\int_{0}^{\infty}\text{d}t\left(1-\frac{\partial}{\partial t}\right)^{n}\left(t^{n-1}e^{-t}\right)=\int_{0}^{\infty}\text{d}t\left(1+\sum_{k=1}^{n}\binom{n}{k}(-1)^{k}\partial_{t}^{k}\right)\left(t^{n-1}e^{-t}\right) \nonumber \\
&=\int_{0}^{\infty}\text{d}t\left(t^{n-1}e^{-t}\right)+\sum_{k=1}^{n}\binom{n}{k}(-1)^{k}\left[\partial_{t}^{k-1}(t^{n-1}e^{-t})\right]_{0}^{\infty}=\Gamma(n)-(-1)^{n}\Gamma(n).
\end{align}


\end{document}